\documentclass[conference]{IEEEtran}

\IEEEoverridecommandlockouts
\usepackage{cite}
\usepackage{amsmath,amsthm,amssymb,array,cases,mathrsfs,subcaption,here,enumitem,ascmac,url,cite,float,comment,color}
\usepackage{algorithmic}
\usepackage{graphicx}
\usepackage{textcomp}
\usepackage{xcolor}
\makeatletter
\let\NAT@parse\undefined
\makeatother
\usepackage{hyperref}

\theoremstyle{plain}
\newtheorem{theorem}{Theorem}
\newtheorem{lemma}{Lemma}
\newtheorem{proposition}{Proposition}

\newtheorem{remark}{Remark}

\theoremstyle{definition}
\newtheorem{definition}{Definition}
\newtheorem{example}{Example}

\renewcommand{\baselinestretch}{0.96}

\usepackage{hyperref}

\begin{document}

\title{\LARGE
On Lie-Algebraic Conditions for Finite-Dimensional \\ Koopman Bilinear Embeddings of Control-Affine Systems
\thanks{The work was partially supported by JSPS KAKENHI (Grant No. 26K00968 and 26KJ1442). 
}
}

\author{Natsuki~Katayama$^1$\thanks{$^1$Natsuki Katayama and Yoshihiko Susuki are with the Department of Electrical Engineering, Kyoto University, Katsura, Nishikyo-ku, Kyoto 615-8510, Japan. E-mails: \texttt{n-katayama@dove.kuee.kyoto-u.ac.jp}, \texttt{susuki.yoshihiko.5c@kyoto-u.ac.jp}} and Yoshihiko~Susuki$^1$}

\maketitle

\begin{abstract}
Koopman bilinear representations for control-affine systems have attracted considerable attention. 
In this paper, we investigate, from a Lie-algebraic perspective, conditions under which a control-affine system admits a finite-dimensional Koopman bilinear embedding.
Specifically, we focus on the Lie algebra generated by the drift and control vector fields. 
First, we show that the finite-dimensionality of the generated Lie algebra is a necessary condition for the existence of a Koopman bilinear embedding.
Next, when the generated Lie algebra contains a nonzero nilpotent ideal, we derive a sufficient condition for the existence of such an embedding.
Finally, through an example of a quadratic polynomial system, we demonstrate that the finite-dimensionality of the generated Lie algebra imposes strong restrictions on the vector fields, suggesting that exact finite-dimensional Koopman bilinear embeddings may be nongeneric.
\end{abstract}

\begin{IEEEkeywords}
Koopman Operator, Nonlinear System, Bilinearization, Lie Algebra
\end{IEEEkeywords}
\IEEEpeerreviewmaketitle

\section{Introduction}

The Koopman operator has recently attracted considerable attention in the analysis and control of nonlinear systems: see, e.g., \cite{mauroy2020koopman,brunton2022modern}.
Originally introduced for nonlinear autonomous systems, it is a linear operator describing the temporal evolution of functions of the state, called observables.
Because the Koopman operator can be approximated from observed data, its data-driven approximation methods, most notably extended dynamic mode decomposition (EDMD), have been widely studied \cite{williams2015data,kutz2016dynamic}.

Koopman-based research on control systems often considers control-affine systems of the form
\begin{equation}
\label{eq:control_affine_intro}
\dot{x}=X_0(x)+\sum_{i=1}^m u_iX_i(x),
\qquad x\in M.
\end{equation}
Early studies investigated Koopman \emph{linear} representations $\dot{z} = Az+Bu$ obtained through a lifting $z=\Psi(x)$ \cite{korda2018linear,proctor2018generalizing}. 
It was subsequently pointed out by \cite{bruder2021advantages} that such a representation need not exist for general control-affine systems. 
Motivated by this, Koopman \emph{bilinear} representations of the form
\begin{equation}
\label{eq:bilinearization_intro}
\dot{z}=A_0z+\sum_{i=1}^m u_iA_iz. 
\end{equation}
have been studied \cite{goswami2021bilinearization,bruder2021advantages}. 
Beyond control-affine systems, more general representations have also been investigated:  see \cite{haseli2023modeling,iacob2024koopman}, the recent review \cite{strasser2026overview}, and references therein. 

The objective of this paper is to establish conditions for the existence of a finite-dimensional embedding $z=\Psi(x)$ that transforms \eqref{eq:control_affine_intro} into the bilinear form \eqref{eq:bilinearization_intro}.
We call such an embedding a \emph{Koopman bilinear embedding}. 
Equivalently, we seek conditions under which the vector fields $X_0,X_1,\ldots,X_m$ admit a common finite-dimensional Koopman invariant subspace (KIS), whose term is coined by  \cite{brunton2016koopman}. 
Establishing these existence conditions has clear benefits. 
For instance, it provides a theoretical basis for assessing whether a finite-dimensional model approximated by EDMD is relevant to an exact finite-dimensional Koopman bilinear embedding.
More fundamentally, it can clarify the nature of 
Koopman representations for control systems, including not only bilinear representations but also infinite-dimensional representations \cite{bevanda2026nonparametric,lazar2026product}, or finite-dimensional models that explicitly account for approximation errors \cite{schaller2023towards,strasser2026safedmd}.

The existence of Koopman embeddings has been investigated in several previous studies.
For nonlinear continuous-time autonomous systems, \cite{liu2025properties,kvalheim2026linearizability} studied the existence of embeddings based on topological properties of the underlying dynamical systems.
For continuous-time systems with a control input $u$, \cite{deka2026differential} characterized a condition under which a Koopman linearization of the form $\dot z=Az+Bu$ exists. 
Koopman linearization for discrete-time systems with a control input has also been studied in \cite{shang2026existence}.
Koopman bilinear embeddings for continuous-time systems have been investigated in \cite{goswami2021bilinearization,bruder2021advantages}.
Their existing conditions for bilinear embeddings require the existence of basis for KIS. 
While finding such a basis is generally challenging \cite{brunton2022modern}, applying these conditions in practice still requires heuristically selecting candidate observables and numerically assessing whether their span is invariant.
Another condition that does not require finding KIS has been expected. 

In this paper, we derive a collection of such conditions for the existence of Koopman bilinear embeddings without finding a basis for KIS. 
Rather than focusing on local differential-geometric properties of the state equations such as \cite{sastry1999nonlinear}, 
we exploit the algebraic structure of vector fields defined globally on $M$. 
Specifically, we use the Lie algebra formed by the vector fields and their Lie brackets. 
The main contributions are the following:
\begin{enumerate}
    \item A Lie-algebraic interpretation for the existence problem of Koopman bilinear embeddings (Proposition~\ref{prop:lie_algebraic_interpretation}).
    \item A necessary condition for the existence of Koopman bilinear embeddings: finite-dimensionality of the Lie algebra generated by $X_0,\ldots,X_m$ (Theorem~\ref{thm:finite_dimensional_generated_lie_algebra}). 
    \item A sufficient condition under which the generated Lie algebra contains a nilpotent ideal (Theorem~\ref{thm:ideal_nilpotent}), together with a comparison with the differential-geometric conditions for Koopman linearizations in \cite{deka2026differential}. 
    \item A quadratic polynomial example demonstrating that the finite-dimensionality of the generated Lie algebra strongly restricts a class of vector fields with Koopman bilinear embedding.
\end{enumerate}
The fourth contribution suggests that the existence of finite-dimensional Koopman bilinear embeddings may be nongeneric, highlighting a strong need to quantify errors in finite-dimensional approximation models.

The remainder of this paper is organized as follows.
Section~\ref{sec:preliminaries} reviews manifolds, Koopman embeddings, Lie algebras, and their representations.
Section~\ref{sec:main_result} derives Lie-algebraic conditions for the existence of Koopman bilinear embeddings.
Section~\ref{sec:example} examines the resulting condition through an example of a quadratic polynomial system.
Section~\ref{sec:conclusion} concludes this paper with a brief summary and future directions.
The proofs of Proposition~\ref{prop:lie_algebraic_interpretation} and Theorem~\ref{thm:ideal_nilpotent} are presented in the appendices.

\section{Preliminaries}
\label{sec:preliminaries}

\subsection{Manifold theory}
\label{subsec:manifold}
This subsection introduces several basic concepts from manifold theory.
Let $M$ be an $n$-dimensional $C^\infty$ manifold. 
For simplicity of coordinate representation, we first consider the case where $M$ is an open subset of $\mathbb{R}^n$.
Let $C^\infty(M;\mathbb{C})$ denote the space of $C^\infty$ complex-valued functions on $M$, and $\mathfrak{X}(M)$ be the space of $C^\infty$ real vector fields on $M$. 
For $X \in \mathfrak{X}(M)$, the \emph{Lie derivative} along $X$ is the linear operator $L_X :C^\infty (M;\mathbb{C}) \to C^\infty (M;\mathbb{C})$ defined by
\[
L_X f (x) := df (x) X(x), \quad f \in C^\infty (M;\mathbb{C}), \quad x\in M,
\]
where $df(x)$ represents the differential of $f$ at $x\in M$. 
In addition, for $X,Y\in \mathfrak{X} (M)$, their \emph{Lie bracket} is the vector field $[X,Y] \in \mathfrak{X}(M)$ defined by 
\begin{equation}
\label{eq:lie_bracket_vector_fields}
[X,Y] (x) := DY (x) X(x) - DX(x) Y(x) ,\quad x \in M    
\end{equation}
where $DX$ and $DY$ represent the Jacobian of $X$ and $Y$, respectively. 
It then follows that $L_{[X,Y]} = L_X L_Y - L_Y L_X$. 
For a general smooth manifold $M$, the Lie derivative and the Lie bracket can be defined by using local coordinates, and the resulting definitions are independent of the choice of local coordinates \cite{Lee_smooth}. 

Next, let $\Psi:M\to \mathbb{C}^N$ be a $C^\infty$ map and $d\Psi(x)$ denote its differential at $x\in M$, where $\mathbb{C}^N$ is regarded as a $2N$-dimensional real manifold. 
The $C^\infty$ map $\Psi$ is called a \emph{$C^\infty$ embedding} if 
$d\Psi(x):T_xM \to T_{\Psi(x)}\mathbb{C}$ is injective for every $x \in M$ and $\Psi:M \to \Psi(M) \subset \mathbb{C}^N$ is a homomorphism. 

\subsection{Koopman embeddings for autonomous systems}
\label{subsec:koopman_autonomous}

This subsection introduces finite-dimensional Koopman linear embeddings for autonomous systems.
Consider the following autonomous system:
\begin{equation}
\label{eq:autonomous_sys}
\dot{x} = X(x) ,\qquad x\in M,
\end{equation}
where $M$ is the state space and $X \in \mathfrak{X}(M)$ is a vector field. 
We simply represent the system \eqref{eq:autonomous_sys} by $(M,X)$.
Throughout this paper, we call a complex-valued function $f \in C^\infty (M;\mathbb{C})$ an observable. 
The Koopman linear embedding problem asks whether a finite collection of observables can be used to embed nonlinear system \eqref{eq:autonomous_sys} into a finite-dimensional linear system. 
\begin{definition}
\label{def:embedding_autonomous}
We say that the autonomous system $(M,X)$ admits a \emph{finite-dimensional Koopman linear embedding} if there exist a positive integer $N$, linearly independent functions $\psi_1,\ldots , \psi_N \in C^\infty (M;\mathbb{C})$, and a matrix $A \in \mathbb{C}^{N\times N}$ such that $\Psi:=(\psi_1,\ldots,\psi_N)^\top:M\hookrightarrow \mathbb{C}^N$ is a $C^\infty$ embedding and satisfies
\[
L_X \Psi := \begin{pmatrix}
    L_X \psi_1 \\ \vdots \\ L_X \psi_N
\end{pmatrix} = A \begin{pmatrix}
    \psi_1 \\ \vdots \\ \psi_N
\end{pmatrix} = A \Psi.
\]
In this case, the pair $(\Psi,A)$ is called a \emph{finite-dimensional Koopman linear embedding} of $(M,X)$, or simply a \emph{Koopman linear embedding}.
\end{definition}
Indeed, if the autonomous system $(M,X)$ admits a Koopman linear embedding $(\Psi,A)$ and $x(t)$ is a trajectory of \eqref{eq:autonomous_sys}, differentiating $z(t) = \Psi (x(t))$ in $t$ yields $\dot{z}(t) = A z(t)$. 
Moreover, because $\Psi$ is an embedding, the state $x$ can be uniquely reconstructed from $z\in\Psi(M)$.
This formulation thus recasts the dynamics from a state-based description, in which one tracks trajectories on $M$, into an observable-based description formulated entirely in terms of observables on $M$ and their Lie derivatives along $X$.

\begin{remark}
Definition~\ref{def:embedding_autonomous} does not require the vector field $X$ to generate a flow globally, although a local flow (trajectory) is always defined. 
If the autonomous system $(M,X)$ generates a semi-flow $\Phi^t:M\to M$ ($t\geq 0$), then the Koopman semi-group can be defined by $U^t f := f \circ \Phi^t$. 
\end{remark}
\begin{remark}
\label{rem:embedding_autonomous_qualitative}
The existence of the Koopman linear embedding for autonomous systems has been studied in the literature \cite{liu2025properties,kvalheim2026linearizability}. 
These studies exploit topological properties of $\omega$-limit sets and attractors to establish the existence of Koopman linear embeddings without explicitly constructing the linear embedding $\Psi$. 
\end{remark}

In Koopman operator theory, the linear operator $L_X:C^\infty (M;\mathbb{C}) \to C^\infty (M;\mathbb{C})$ is called the Koopman generator. 
Note that if the Koopman semi-group exists for $t\geq 0$, then $L_X$ is identical to its infinitesimal generator. 
By Definition~\ref{def:embedding_autonomous}, a finite-dimensional subspace $\mathcal{F} := \operatorname{span}_\mathbb{C} \{ \psi_1, \ldots, \psi_N \} \subset C^\infty (M;\mathbb{C})$ satisfies $L_X \mathcal{F} \subseteq \mathcal{F}$. 
Such a subspace $\mathcal{F}$ is called a Koopman invariant subspace (KIS). 
In particular, we have a nice spectral property as $\sigma(L_X|_{\mathcal{F}}) = \sigma(A)$, where $\sigma (L_X|_{\mathcal{F}})$ and $\sigma (A)$ represent the set of eigenvalues of $L_X|_\mathcal{F}$ and $A$, respectively. 

\subsection{Lie algebras and representations}
\label{subsec:lie_algebra_representation}
This subsection introduces the basic concepts of Lie algebras and their representations \cite{Humphreys1972LieAlgebras,elliott2009bilinear}. 
In what follows, Lie algebras are not assumed to be finite-dimensional.
\begin{definition}
A real vector space $\mathfrak{g}$, equipped with a bilinear map $[\cdot , \cdot] : \mathfrak{g} \times \mathfrak{g} \to \mathfrak{g}$,
is called a \emph{real Lie algebra} (or simply a \emph{Lie algebra}) if the following conditions hold:
\begin{enumerate}
    \item $[X,X] = 0$ for every $X\in \mathfrak{g}$;
    \item $[X,[Y,Z]] + [Y,[Z,X]] + [Z,[X,Y]] = 0$ for every $X,Y,Z\in \mathfrak{g}$ (\emph{Jacobi identity}).
\end{enumerate}
The bilinear map $[\cdot , \cdot]$ is called the \emph{Lie bracket} on $\mathfrak{g}$.
\end{definition}
\begin{example}
The space $\mathfrak{X} (M)$ of smooth vector fields on $M$, equipped with the Lie bracket \eqref{eq:lie_bracket_vector_fields}, is an infinite-dimensional real Lie algebra. 
\end{example}
\begin{example}
Let $V$ be a complex vector space, and let $\operatorname{End}(V)$ be the space of all complex-linear endomorphisms of $V$.
Regarding $\operatorname{End}(V)$ as a real vector space, we define 
\[
[A,B]_{\mathrm{End}} v := A (Bv) - B (Av),\quad A,B\in\operatorname{End}(V), \ v\in V.
\]
Then, $\operatorname{End}(V)$, equipped with this bracket, is a real Lie algebra. 
\end{example}

To develop the main results of this paper, we introduce the following notions:
\begin{definition}
\label{def:subalgebra}
A vector subspace $\mathfrak{h}\subseteq \mathfrak{g}$ is called a \emph{Lie subalgebra} if $h_1,h_2\in \mathfrak{h} \implies [h_1,h_2]\in \mathfrak{h}$. 
\end{definition}
\begin{definition}
\label{def:generated_lie_algebra}
Let $h_1,\ldots,h_m \in \mathfrak{g}$. 
The smallest Lie subalgebra of $\mathfrak{g}$ containing $h_1,\ldots,h_m$ is called \emph{Lie algebra generated by $h_1,\ldots,h_m$} and denoted by $\operatorname{Lie}(h_1,\ldots,h_m)$. 
Equivalently, we have 
\[
\begin{aligned}
\operatorname{Lie}(h_1,\ldots,h_m) = &\operatorname{span}_\mathbb{R} \bigl\{ h_i, [h_i,h_j], [h_i,[h_j,h_k]], \\
&\quad [[h_i,h_j],h_k], [[h_i,h_j],[h_k,h_\ell]], \ldots \bigr\},
\end{aligned}
\]
where the indices range over $\{1,\ldots,m\}$, and all possible finite iterated Lie brackets are included.
In particular, if $\mathfrak{g}$ is infinite-dimensional, $\operatorname{Lie}(h_1,\ldots,h_m)$ may be infinite-dimensional even when $m$ is finite.
\end{definition}

Next, we introduce a representation of a Lie algebra. 
\begin{definition}
\label{def:representation}
Let $\mathfrak{g}$ be a real Lie algebra and $V$ be a complex vector space. 
A \emph{representation of $\mathfrak{g}$ on $V$} is a real linear map $\pi:\mathfrak{g} \to \operatorname{End} (V)$ such that
\begin{equation}
\label{eq:representation_homomorphism}
\pi([X,Y]) = [\pi(X), \pi(Y)]_{\rm End}, \qquad \forall X,Y\in \mathfrak{g}, 
\end{equation}
where $\operatorname{End} (V)$ is regarded as a real Lie algebra. 
\end{definition}
The Lie derivative provides a representation of the Lie algebra $\mathfrak{X}(M)$: 
\begin{proposition}
\label{prop:representation_lie_derivative}
For Definition~\ref{def:representation}, set $\mathfrak{g} = \mathfrak{X} (M)$ and $V = C^\infty (M;\mathbb{C})$. 
Then, 
\[
\pi: \mathfrak{X}(M) \to \operatorname{End} (C^\infty (M;\mathbb{C})), \quad \pi(X):= L_X
\]
is a representation of $\mathfrak{X}(M)$.
\end{proposition}
\begin{proof}
The map $\pi$ is real linear because $\pi(aX+bY) = L_{aX + bY} = a L_X + b L_Y$ for every $a,b\in \mathbb{R}$ and every $X,Y \in \mathfrak{X} (M)$. 
Equation~\eqref{eq:representation_homomorphism} follows from
$\pi([X,Y])= L_{[X,Y]} = L_X L_Y - L_Y L_X = [\pi(X),\pi(Y)]_{\operatorname{End}}$. 
\end{proof}

\section{Main results: Lie-algebraic conditions for Koopman bilinear embeddings}
\label{sec:main_result}
This section presents our main results on Lie-algebraic conditions for Koopman bilinear embeddings. 
Section~\ref{subsec:problem_statement} formulates the problem; Section~\ref{subsec:lie_algebra_interpretation} gives its Lie-algebraic interpretation; Section~\ref{subsec:necessary_condition} derives a necessary condition; and Section~\ref{subsec:sufficient_condition} establishes a sufficient condition involving a nilpotent ideal and compares it with the existing conditions of Koopman linearization in \cite{deka2026differential}.

\subsection{Problem statement}
\label{subsec:problem_statement}
This subsection formulates the existence problem for finite-dimensional Koopman bilinear embeddings of control-affine systems. 
Consider the following control-affine system:
\begin{equation}
\label{eq:control_affine}
\dot x = X_0 (x) + \sum_{j=1}^m u_j (t) X_j(x), \qquad x\in M,
\end{equation}
where $M$ is an $n$-dimensional $C^\infty$ manifold, $X_0,\ldots,X_m \in \mathfrak{X} (M)$ are $C^\infty$ vector fields, and $u=(u_1,\ldots,u_m)^\top \in \mathbb{R}^m$ is the control input. 
We simply represent the system \eqref{eq:control_affine} by $(M,X_j)_{0\leq j\leq m}$.
The existence problem for a Koopman bilinear embedding is to pose the question of whether a finite collection of observables can be used to embed the control-affine system \eqref{eq:control_affine} into a finite-dimensional bilinear system of the form
\begin{equation}
\label{eq:bilinearized_sys}
\dot{z} = A_0 z + \sum_{j=1}^m u_j (t) A_j z, \qquad  z\in \mathbb{C}^N. 
\end{equation}
\begin{definition}
\label{def:embedding_bilinear}
We say that the control-affine system $(M,X_j)_{0\leq j \leq m}$ admits a \emph{finite-dimensional Koopman bilinear embedding} if there exist a positive integer $N$, linearly independent functions $\psi_1,\ldots, \psi_N \in C^\infty (M;\mathbb{C})$, and matrices $A_0,\ldots,A_m\in \mathbb{C}^{N\times N}$ such that $\Psi:= (\psi_1,\ldots, \psi_N )^\top:M\hookrightarrow \mathbb{C}^N$ is a $C^\infty$ embedding and satisfies
\begin{equation}
\label{eq:lie_derivative_bilinear}
L_{X_j} \Psi = A_j \Psi,\qquad  j=0,\ldots,m.
\end{equation}
In this case, the pair $(\Psi,A_j)_{0\leq j \leq m}$ is called a \emph{finite-dimensional Koopman bilinear embedding} of $(M,X_j)_{0\leq j \leq m}$, or simply a \emph{Koopman bilinear embedding}. 
\end{definition}
Indeed, if there exists a Koopman bilinear embedding $(\Psi,A_j)_{0\leq j \leq m}$, then differentiating $z (t)= \Psi (x(t))$ in $t$ yields \eqref{eq:bilinearized_sys}. 
The existence of matrices $A_0,\ldots,A_m$ satisfying \eqref{eq:lie_derivative_bilinear} is equivalent to the invariance of the finite-dimensional subspace $\mathcal{F} := \operatorname{span}_\mathbb{C} \{\psi_1,\ldots, \psi_N\}$ under all the Lie derivatives $L_{X_j}$. 
Namely, $\mathcal{F}$ is a \emph{common} KIS: 
\begin{equation}
\label{eq:Koopman_invariant_subspace_control_affine}
L_{X_j} \mathcal{F} \subseteq \mathcal{F} ,\qquad  j=0,\ldots,m.
\end{equation}
Definition~\ref{def:embedding_autonomous} is applicable to each $X_j$ and requires the existence of a KIS for each $X_j$, while Definition~\ref{def:embedding_bilinear} requires the existence of a common KIS for all the vector fields, that is, a \emph{common} embedding. 

\begin{remark}
\label{rem:bilinear_include_linear}
The constant function $1$ may be included as a component of $\Psi$.
{\color{black}Thus, systems of the form
\begin{equation}
\label{eq:bilinear_linear}
\dot{z} = A_0 z + b_0 + \sum_{j=1}^m u_j(t) (A_jz + b_j),
\end{equation}
including linear control systems,}
can be represented as bilinear systems: see \cite{bruder2021advantages}. 
\end{remark}

The main difficulty in establishing the existence of a finite-dimensional Koopman bilinear embedding is that Definition~\ref{def:embedding_bilinear} requires a single embedding map $\Psi$ that is common to all the vector fields $X_0,\ldots,X_m$. 
We will resolve this difficulty by investigating an algebraic structure of the vector fields.

\subsection{Lie-algebraic interpretation of Koopman bilinear embeddings}
\label{subsec:lie_algebra_interpretation}
This subsection provides a Lie-algebraic interpretation of the existence problem for Koopman bilinear embeddings. 
Recall from Proposition~\ref{prop:representation_lie_derivative} that $\mathfrak{X}(M)$ is a real Lie algebra and that $\pi(X) := L_X$ is a representation. 
Let
\[
\mathfrak{Y} := \operatorname{Lie} (X_0,\ldots, X_m)
\]
denote the Lie subalgebra\footnote{A Lie algebra of vector fields should be distinguished from a distribution, which is a subbundle of the tangent bundle $TM$. 
} of $\mathfrak{X}(M)$ generated by $X_0,\ldots,X_m$ as defined in Definition~\ref{def:generated_lie_algebra}.
This Lie algebra $\mathfrak{Y}$ may be infinite-dimensional even though it is generated by finitely many vector fields. 
The following proposition provides a Lie-algebraic interpretation of the existence problem for Koopman bilinear embedding. 
\begin{proposition}
\label{prop:lie_algebraic_interpretation}
The control-affine system $(M,X_j)_{0\leq j\leq m}$ admits a Koopman bilinear embedding if and only if there exists a finite-dimensional subspace $\mathcal{F} \subset C^\infty (M;\mathbb{C})$ with the following two properties:
\begin{enumerate}
    \item[\textup{(i)}] For some basis $\psi_1,\ldots,\psi_N$ of $\mathcal{F}$, the map $\Psi=(\psi_1,\ldots,\psi_N)^\top : M \to \mathbb{C}^N$ is a $C^\infty$ embedding;
    \item[\textup{(ii)}] 
    The representation
    \begin{equation}
    \label{eq:pi_Y_def}
    \pi_\mathfrak{Y} : \mathfrak{Y} \to \operatorname{End} (\mathcal{F}) ,\qquad \pi_\mathfrak{Y}(Y) := L_Y |_{\mathcal{F}}
    \end{equation}
    is well-defined and injective. 
\end{enumerate}
\end{proposition}
\begin{proof}
See Appendix~\ref{sec:proof_lie_algebraic_interpretation}. 
\end{proof}

Property~(i) ensures that a basis of $\mathcal{F}$ defines an embedding of $M$. 
The well-definedness of $\pi_\mathfrak{Y}$ implies 
\begin{equation}
\label{eq:invariance_every_Y}
L_Y \mathcal{F} \subseteq \mathcal{F} ,\quad \forall Y \in \mathfrak{Y}. 
\end{equation}
Property~\textup{(ii)} thus reformulates the common-invariance condition \eqref{eq:Koopman_invariant_subspace_control_affine} as an invariance condition under $L_Y$ for every $Y\in\mathfrak{Y}$, including not only $X_0,\ldots,X_m$ but also their Lie brackets $[X_i,X_j]$ and iterated Lie brackets $[[X_i,X_j],X_k], [[X_i,X_j],[X_k,X_\ell]], \ldots$ 
These additional vector fields enable us to unravel a problem, for instance, to describe a structure that is never clarified with the original vector fields only.
The Lie-algebraic interpretation in Proposition~\ref{prop:lie_algebraic_interpretation} leads to the necessary condition in the next subsection.

\subsection{A necessary condition}
\label{subsec:necessary_condition}
We now derive a necessary condition for the existence of a Koopman bilinear embedding based on Proposition~\ref{prop:lie_algebraic_interpretation}.  
Property~(ii) 
yields the following necessary condition. 
\begin{theorem}
\label{thm:finite_dimensional_generated_lie_algebra}
If the control-affine system $(M,X_j)_{0\leq j\leq m}$ admits a Koopman bilinear embedding $(\Psi,A_j)_{0\leq j \leq m}$, then the following condition holds: 
\begin{enumerate}
    \item[\textup{(I)}] The generated Lie algebra $\mathfrak{Y} = \operatorname{Lie}(X_0,\ldots, X_m)$ is finite-dimensional. 
\end{enumerate}
\end{theorem}
\begin{proof}
Since $\pi_\mathfrak{Y}:\mathfrak{Y} \to \operatorname{End} (\mathcal{F})$ is injective, $\dim \mathfrak{Y} \leq \dim \operatorname{End} (\mathcal{F})$. 
Thus $\dim \mathcal{F} < \infty$ implies $\dim \mathfrak{Y} \leq \dim \operatorname{End} (\mathcal{F})<\infty$. 
\end{proof}
Thus, Condition~(I) requires the Lie algebra $\mathfrak{Y}$ generated by $X_0,\ldots,X_m$ to be finite-dimensional. 
Since $\mathfrak{Y}$ contains not only $X_0,\ldots,X_m$ but also their Lie brackets 
and iterated Lie brackets, 
the finite-dimensionality in Condition~(I) imposes a structural restriction on the class of systems that admit a common embedding $\Psi$.

It should be stated that Condition~(I) can be investigated directly from the vector fields, without explicitly constructing $\Psi$.
More precisely, define a sequence of real vector subspaces of $\mathfrak{X}(M)$ by
\begin{equation}
\label{eq:lie_sequence}
\begin{aligned}
\mathfrak{Y}^{(1)} &= \operatorname{span}_\mathbb{R} \{X_0,\ldots,X_m \} \\
\mathfrak{Y}^{(k+1)} &= \mathfrak{Y}^{(k)} \\
& + \operatorname{span}_\mathbb{R} \{ [X_j, Y] \; | \; 0\leq j \leq m, \; Y \in \mathfrak{Y}^{(k)} \},
\end{aligned}
\end{equation}
where $k\geq 1$.
Each $\mathfrak{Y}^{(k)}$ is clearly finite-dimensional, and we have\footnote{It is not immediate that the generated Lie algebra $\mathfrak{Y}$ coincides with $\bigcup_{k=1}^{\infty}\mathfrak{Y}^{(k)}$. One way to prove this is to use a \emph{Philip Hall basis}; see, for example, \cite[Appendix~B]{elliott2009bilinear} and \cite[p.~518]{sastry1999nonlinear}.}
\begin{equation}
\nonumber
\mathfrak{Y}^{(1)} \subseteq \mathfrak{Y}^{(2)} \subseteq \cdots,\quad \mathfrak{Y} = \textstyle\bigcup_{k=1}^{\infty}\mathfrak{Y}^{(k)}.
\end{equation}
If there exists an integer $k\geq 1$ such that $\mathfrak{Y}^{(k+1)} = \mathfrak{Y}^{(k)}$, then we see that a finite-dimensional generated Lie algebra $\mathfrak{Y} = \mathfrak{Y}^{(k)}$ exists. 
If no such $k$ exists, then the dimensions of $\mathfrak{Y}^{(k)}$ increase without bound, that is to say, $\mathfrak{Y}$ is infinite-dimensional. 
In such a case, by Theorem~\ref{thm:finite_dimensional_generated_lie_algebra}, we conclude that no Koopman bilinear embedding exists for the control-affine system.

\subsection{A sufficient condition via nilpotent ideals}
\label{subsec:sufficient_condition}

To derive a sufficient condition for Koopman bilinear embedding, we investigate the structure of the Lie algebra $\mathfrak{Y} = \operatorname{Lie}(X_0,\ldots,X_m)$. 
Specifically, in this subsection, we derive a sufficient condition based on the existence of a nilpotent ideal in $\mathfrak{Y}$. 

We first introduce the notion of a finite-algebraic-multiplicity Koopman linear embedding. 
Let $L_X:C^\infty(M;\mathbb{C}) \to C^\infty(M;\mathbb{C})$ be the Koopman generator associated with a vector field $X\in \mathfrak{X} (M)$. 
For $\lambda \in \mathbb{C}$, using the kernel space of the operator $L_X-\lambda I$ ($I$ is the identity), define
\[
\mathcal{R}_\lambda (L_X) := \textstyle \bigcup_{k=1}^\infty\operatorname{Ker} (L_X - \lambda I)^k
\]
as a subspace of $C^\infty(M;\mathbb{C})$. 
This is called the \emph{algebraic root space}\footnote{If $\dim \mathcal{R}_\lambda(L_X)<\infty$, then this space corresponds to the generalized eigenspace of $L_X$ associated with $\lambda$. } of $L_X$ associated with $\lambda$. 
Then, the finite-algebraic-multiplicity Koopman linear embedding is defined as follows.
\begin{definition}
\label{def:finite_algebraic_multiplicity}
A Koopman linear embedding $(\Psi,A)$ of the autonomous system $(M,X)$ is said to have \emph{finite algebraic multiplicity} if $\dim \mathcal{R}_\lambda(L_X)<\infty$ for every $\lambda\in\sigma(A)$. 
We also say that $(M,X)$ admits a \emph{finite-algebraic-multiplicity Koopman linear embedding}.
\end{definition}
As a simple example, consider the system $\dot{x}=\operatorname{diag}(1,a)x$ with $x=(x_1,x_2)^\top$. 
For $a>0$, the identity map itself is a finite-algebraic-multiplicity Koopman linear embedding. 
If $a=0$, however, $x_2^k\in\mathcal{R}_0(L_X)$ for every $k\geq0$, so $\dim\mathcal{R}_0(L_X)=\infty$ and a finite-algebraic-multiplicity Koopman linear embedding does not exist. 

Next, we recall the notions of an ideal and a nilpotent Lie algebra from \cite{Humphreys1972LieAlgebras}. 
\begin{definition}
A Lie subalgebra $\mathfrak{h}\subset \mathfrak{g}$ is called an \emph{ideal} if $[X,Z] \in \mathfrak{h}$ for every $X\in \mathfrak{g}$ and $Z\in \mathfrak{h}$. 
\end{definition}
\begin{definition}
\label{def:nilpotent}
A Lie algebra $\mathfrak{g}$ is called \emph{nilpotent} if there exists a positive integer $r$ such that $[Y_1,[Y_2,\ldots,[Y_r,Y_{r+1}]\ldots]]=0$ for all $Y_1,\ldots,Y_{r+1}\in\mathfrak{g}$.
\end{definition}

With these definitions, we have the following sufficient condition.
\begin{theorem}
\label{thm:ideal_nilpotent}
The control-affine system $(M,X_j)_{0\leq j \leq m}$ admits a Koopman bilinear embedding $(\Psi,A_j)_{0\leq j \leq m}$ if the following conditions hold:
\begin{enumerate}
    \item[\textup{(A)}] The generated Lie algebra $\mathfrak{Y} = \operatorname{Lie}(X_0,\ldots,X_m)$ is finite-dimensional and contains a nonzero nilpotent ideal $\mathfrak{I}$.
    \item[\textup{(B)}] There exists a vector field $Z \in \mathfrak{I}$ such that the autonomous system $(M,Z)$ admits a finite-algebraic-multiplicity Koopman linear embedding. 
\end{enumerate}
\end{theorem}
\begin{proof}
    See Appendix~\ref{sec:proof_ideal_nilpotent}. 
\end{proof}
As a simple example, suppose that $X_0,\ldots,X_m$ commute pairwise. 
Then, $\mathfrak{I}=\operatorname{span}_\mathbb{R} \{ X_0,\ldots,X_m \}$ is a commutative Lie subalgebra and hence is nilpotent, and Condition~(A) holds with $\mathfrak{I} = \mathfrak{Y}$. 
In this case, the control-affine system $(M,X_j)_{0\leq j \leq m}$ admits a Koopman bilinear embedding if one of the autonomous systems $(M,X_0),\ldots,(M,X_m)$ admits a finite-algebraic-multiplicity Koopman linear embedding in Definition~\ref{def:finite_algebraic_multiplicity}. 

It should be noted that not every finite-dimensional Lie algebra contains a nonzero nilpotent ideal. 
The conditions in Theorem~\ref{thm:ideal_nilpotent} are therefore restrictive.
Establishing a sufficient condition applicable to more general classes of generated Lie algebras is left for future work. 

\begin{remark}
\label{rem:comparison}
We compare Theorem~\ref{thm:ideal_nilpotent} with the differential-geometric conditions for Koopman linear embedding reported in \cite{deka2026differential}.
That work considers a local map $z=\Phi(x)$, defined on a neighborhood of an equilibrium point $x_{0}$ of the drift vector field $X_0$, that transforms a control-affine system into a linear system of the form $\dot{z} = Az + Bu$. 
By taking $M$ to be a sufficiently small neighborhood of $x_{0}$, \cite{deka2026differential} shows that such a transformation into a controllable linear system exists if and only if the following conditions hold: 
\begin{enumerate}
    \item[\textup{(a1)}] $\Delta (x) := \bigl\{ \operatorname{ad}_{X_0}^k X_i (x) \, \big| \,1 \leq i\leq m,\, k\geq0 \bigr\}$ satisfies $\Delta(x)=T_xM$ for every $x\in M$.
    \item[\textup{(a2)}] $\bigl[ \operatorname{ad}_{X_0}^kX_i, \operatorname{ad}_{X_0}^{\ell}X_j \bigr] = 0$ for every $1\leq i,j\leq m$ and $k,\ell\geq0$. 
\end{enumerate}
Here, the iterated adjoint actions are defined recursively by $\operatorname{ad}_X^0Y:=Y$ and $\operatorname{ad}_X^kY:=[X,\operatorname{ad}_X^{k-1}Y]$ for $k\geq 0$. 
Although these conditions are formulated locally, they become necessary conditions for a global Koopman linearization of the form $\dot{z}=Az+Bu$. 

To compare Conditions~\textup{(a1)} and \textup{(a2)} with Theorem~\ref{thm:ideal_nilpotent}, define 
\[
\mathfrak{I}_{\mathrm{D}} := \operatorname{span}_{\mathbb{R}} \bigl\{ \operatorname{ad}_{X_0}^kX_i \,\big|\, 1\leq i\leq m,\quad k\geq0 \bigr\}.
\]
Condition~\textup{(a2)} implies that $\mathfrak{I}_{\mathrm{D}}$ is a commutative Lie algebra containing $X_1,\ldots,X_m$. 
Condition~\textup{(a2)} also implies that
$
[X_0,Y] = \operatorname{ad}_{X_0} Y \in \mathfrak{I}_{\mathrm{D}}$ for every $Y \in \mathfrak{I}_{\mathrm{D}}. 
$
Thus, $\mathfrak{I}_{\mathrm{D}}$ is an ideal of $\mathfrak{Y} = \operatorname{Lie}(X_0,\ldots,X_m)$ and 
\begin{equation}
\label{eq:generated_lie_algebra_decomposition}
    \mathfrak{Y} = \operatorname{span}_\mathbb{R} \{X_0\} + \mathfrak{I}_{\mathrm{D}}.
\end{equation}
Furthermore, Condition~\textup{(a1)} implies that $\mathfrak{I}_{\mathrm{D}}$ is transitive\footnote{A Lie algebra $\mathfrak{g}$ of vector fields is said to be \emph{transitive} if $\operatorname{span}_\mathbb{R}\{ X(x) \,|\, X\in \mathfrak{g} \} = T_xM$ for every $x\in M$. }, which, together with commutativity of $\mathfrak{I}_{\mathrm{D}}$, yields $\dim \mathfrak{I}_{\mathrm{D}} = n$; see \cite[pp.231--236]{Lee_smooth}. 
Consequently, we have $\dim \mathfrak{Y} \leq n+1 < \infty$. 
Therefore, Conditions~\textup{(a1)} and \textup{(a2)} imply Condition~\textup{(A)} of Theorem~\ref{thm:ideal_nilpotent}. 

From this Lie-algebraic viewpoint, we can state that the structure arising from the two conditions in \cite{deka2026differential} is a special case of Condition~\textup{(A)} in which the nilpotent ideal $\mathfrak{I}$ is commutative and the generated Lie algebra takes the form \eqref{eq:generated_lie_algebra_decomposition}. 
In this sense, the Lie-algebraic requirement in Condition~\textup{(A)} is more general than the structure imposed by Conditions~\textup{(a1)} and \textup{(a2)}. 
Since a linear system of the form $\dot z = Az + Bu$ can be represented as a bilinear system as stated in Remark~\ref{rem:bilinear_include_linear}, the linear systems considered in \cite{deka2026differential} are included in our bilinear formulation. 

However, Theorem~\ref{thm:ideal_nilpotent} guarantees only the existence of a Koopman bilinear embedding and does not characterize when such an embedding can be reduced to a linear system of the form $\dot z = Az + Bu$. 
It also remains unclear whether Conditions~\textup{(a1)} and \textup{(a2)} imply Condition~\textup{(B)}. 
Clarifying the relationship between these conditions and Condition \textup{(B)}, particularly from a differential-geometric viewpoint, is left for future work. 
\end{remark}

\section{Example and discussion}
\label{sec:example}
This subsection illustrates how Condition~(I) in Theorem~\ref{thm:finite_dimensional_generated_lie_algebra} can be checked for a two-dimensional, single-input polynomial system. 
Let $M=\mathbb{R}^2$, $m=1$, and consider 
\begin{equation}
\label{eq:example}
\begin{pmatrix}
    \dot{x} \\ \dot{y}
\end{pmatrix} = \begin{pmatrix}
    y \\ f(x,y)
\end{pmatrix} + u \begin{pmatrix}
    0 \\ 1
\end{pmatrix},
\end{equation}
where $f:\mathbb{R}^2 \to \mathbb{R}$ is a real polynomial of degree at most two. 
Equivalently, consider the vector fields
\[
X_0(x,y) := y \frac{\partial}{\partial x} + f(x,y) \frac{\partial}{\partial y}, \quad X_1(x,y) := \frac{\partial}{\partial y}. 
\]
This system can be viewed as a quadratic Taylor approximation at the origin of a one-degree-of-freedom equation of motion $\ddot{x}=F(x,\dot{x})+u$ with $F(x,y)=f(x,y)+O(\|(x,y)\|^3)$ where $O$ represents Landau's big-$O$ notation. 

The following proposition characterizes the polynomials $f$ for which the generated Lie algebra is finite-dimensional. 
\begin{proposition}
\label{prop:example_condition}
For the system \eqref{eq:example}, the generated Lie algebra $\operatorname{Lie}(X_0,X_1)$ is finite-dimensional if and only if 
\begin{equation}
\label{eq:example_condition}
\frac{\partial^2 f}{\partial x^2} (0) = \frac{\partial^2 f}{\partial x \partial y} (0) = \frac{\partial^2 f}{\partial y^2} (0) \frac{\partial f}{\partial x} (0) = 0.
\end{equation}
Equivalently, $f$ has one of the forms of 
\[
f(x,y) = f_{02} y^2 + f_{01} y + f_{00} ~~ \text{or} ~~ f(x,y) = f_{10} x + f_{01} y + f_{00}
\]
where $f_{02} , f_{10}, f_{01}, f_{00}\in \mathbb{R}$. 
\end{proposition}

A complete proof of Proposition~\ref{prop:example_condition} covering all
possible coefficients is omitted. 
Instead, Table~\ref{tab:calculation} compares two representative cases: $f(x,y)=y^2+y+1$, which satisfies~\eqref{eq:example_condition}, and $f(x,y)=xy$, which violates it.
For $f(x,y)=y^2+y+1$, Table~\ref{tab:calculation} shows that $\mathfrak{Y}^{(4)}=\mathfrak{Y}^{(3)}$. 
As demonstrated in Section~\ref{subsec:necessary_condition}, this implies that $\operatorname{Lie}(X_0,X_1)=\mathfrak{Y}^{(3)}$, which is four-dimensional.
For $f(x,y)=xy$, by contrast, the table shows polynomial vector fields of progressively higher degree. 
Proposition~\ref{prop:example_condition} establishes that $\operatorname{Lie}(X_0,X_1)$ is infinite-dimensional in this case. 
Similar degree growth occurs for other polynomials that violate \eqref{eq:example_condition}. 
These calculations illustrate that the sequence~\eqref{eq:lie_sequence} provides a useful diagnostic for the
finite-dimensionality of the generated Lie algebra. 
\begin{table}[t]
    \vspace*{4.5pt}
    \caption{Sequence~\eqref{eq:lie_sequence} for two representative polynomials. Here, the notation $\langle v_1,\ldots,v_r\rangle:=\operatorname{span}\{v_1,\ldots,v_r\}$ is used.}
    \centering
    \small
    \setlength{\tabcolsep}{5pt}
    \renewcommand{\arraystretch}{1.8}
    \label{tab:calculation}
    \begin{tabular}{c|l|l}
      \hline
      & $f(x,y)=y^2+y+1$ & $f(x,y)=xy$ \\
      \hline
      $\mathfrak{Y}^{(1)}$
      & \scalebox{1.2}{$\bigl\langle \binom{0}{1},\binom{y}{y^2+y+1}\bigr\rangle$}
      & \scalebox{1.2}{$\bigl\langle \binom{0}{1},\binom{y}{xy}\bigr\rangle$} \\
      $\mathfrak{Y}^{(2)}$
      & $\mathfrak{Y}^{(1)}+{}$\scalebox{1.2}{$\bigl\langle\binom{1}{2y+1}\bigr\rangle$}
      & $\mathfrak{Y}^{(1)}+{}$\scalebox{1.2}{$\bigl\langle\binom{1}{x}\bigr\rangle$} \\
      $\mathfrak{Y}^{(3)}$
      & $\mathfrak{Y}^{(2)}+{}$\scalebox{1.2}{$\bigl\langle\binom{2y+1}{2y^2+2y-1}\bigr\rangle$}
      & $\mathfrak{Y}^{(2)}+{}$\scalebox{1.2}{$\bigl\langle\binom{x}{x^2}\bigr\rangle$} \\
      $\mathfrak{Y}^{(4)}$
      & $\mathfrak{Y}^{(3)}$
      & $\mathfrak{Y}^{(3)}+{}$\scalebox{1.2}{$\bigl\langle\binom{x^2-y}{x^3-xy}\bigr\rangle$} \\
      $\mathfrak{Y}^{(5)}$
      & $\mathfrak{Y}^{(4)}$
      & $\mathfrak{Y}^{(4)}+{}$\scalebox{1.2}{$\bigl\langle\binom{x^3-2xy}{x^4-2x^2y}\bigr\rangle$} \\
      \hline
    \end{tabular}
\end{table}

Proposition~\ref{prop:example_condition} can be interpreted as follows:
Let $\mathcal{P}^{\leq2}(\mathbb R^2;\mathbb R)$ represent the vector space of real polynomials in $(x,y)$ of degree at most two. 
Because $f$ can be  written as
$
f(x,y) = f_{20}x^2+f_{11}xy+f_{02}y^2+f_{10}x+f_{01}y+f_{00},
$
$\mathcal{P}^{\leq2}(\mathbb R^2;\mathbb R)$ can be identified with $\mathbb R^6$. 
Define 
\[
\mathcal{P}_{\mathrm{finite}}^{\leq2} := \{ f\in\mathcal{P}^{\leq2}(\mathbb R^2;\mathbb R) \,|\, \dim\operatorname{Lie}(X_0,X_1)<\infty\}.
\]
By Proposition~\ref{prop:example_condition}, we derive
\[
\begin{aligned}
\mathcal{P}_{\mathrm{finite}}^{\leq2} ={}& \left\{ f\in\mathcal{P}^{\leq2}(\mathbb R^2;\mathbb R) \, \middle| \, f_{20}=f_{11}=f_{02}=0 \right\}\\
&\quad\cup \left\{ f\in\mathcal{P}^{\leq2}(\mathbb R^2;\mathbb R) \, \middle| \, f_{20}=f_{11}=f_{10}=0 \right\}.
\end{aligned}
\]
Thus, $\mathcal{P}_{\mathrm{finite}}^{\leq2}$ is 
nowhere dense and has Lebesgue measure zero with respect to $\mathcal{P}^{\leq 2}(\mathbb{R}^2,\mathbb{R})$.
Hence, 
Theorem~\ref{thm:finite_dimensional_generated_lie_algebra} implies that the set of polynomials $f$ for which a finite-dimensional Koopman bilinear embedding can exist 
is nowhere dense and of Lebesgue measure zero. 
In this sense, an exact finite-dimensional Koopman bilinear embedding is {nongeneric} for $f\in \mathcal{P}^{\leq2}(\mathbb R^2;\mathbb R)$. 

This conclusion is restricted to two-dimensional, single-input systems with quadratic polynomial dynamics of the form \eqref{eq:example} and does not directly extend to general control-affine systems. 
Nevertheless, this example demonstrates that the finite-dimensionality of the generated Lie algebra imposes a strong algebraic restriction. 
It therefore suggests that one should be cautious about assuming the existence of an exact finite-dimensional Koopman bilinear embedding without additional structure.

This observation also has implications for data-driven Koopman modeling. 
In general, a learned finite-dimensional model should be regarded as an approximation of the form
\[
\dot z = A_0z+\sum_{j=1}^m u_jA_jz+r_\Psi(z,u),
\]
where $r_\Psi(z,u)$ accounts for the part of the lifted dynamics that cannot be captured within the finite-dimensional span of the components of $\Psi$.
Thus, even when a large dictionary of observables is employed, it is important to assess the residual $r_\Psi(z,u)$ explicitly; see \cite{schaller2023towards,strasser2026safedmd} and \cite[Section~4]{strasser2026overview}. 

\section{Conclusion}
\label{sec:conclusion}

This paper investigated the existence of finite-dimensional Koopman bilinear embeddings for control-affine systems from the perspective of the Lie algebra generated by the underlying vector fields and its representations.
We showed that the finite-dimensionality of the generated Lie algebra is a necessary condition for such an embedding. 
We also derived a sufficient condition applicable when the generated Lie algebra contains a nilpotent ideal.
Our analysis of a quadratic polynomial system indicates that the finite-dimensionality imposes strong restrictions on a class of vector fields, suggesting that finite-dimensional Koopman bilinear embeddings may be nongeneric.
Future work will introduce appropriate norms on the generated Lie algebra and develop a quantitative theory of finite-dimensional bilinear approximations and their approximation errors.

\section*{Acknowledgement}
The authors used ChatGPT (OpenAI) to assist with the derivations of Theorem~\ref{thm:ideal_nilpotent} and Proposition~\ref{prop:example_condition}, the comparison of Theorem~\ref{thm:ideal_nilpotent} with the result reported in \cite{deka2026differential}, the identification and correction of issues in the proofs of Proposition~\ref{prop:lie_algebraic_interpretation} and Theorem~\ref{thm:ideal_nilpotent}, the calculations in Section~\ref{sec:example}, searches for relevant literature, and language editing to improve the clarity and readability of the manuscript. 
All AI-generated suggestions were reviewed and edited by the authors to ensure accuracy and integrity.

\appendices
\section{Proof of Proposition~\ref{prop:lie_algebraic_interpretation}}
\label{sec:proof_lie_algebraic_interpretation}
\begin{proof}
$(\Longrightarrow)$ Assume that there exist a Koopman bilinear embedding $(\Psi,A_j)_{0\leq j\leq m}$. 
Set $\mathcal{F} = \operatorname{span}_\mathbb{C} \{ \psi_1,\ldots, \psi_N\}$. 
Property~(i) follows directly from the definition of a Koopman bilinear embedding. 
{\color{black}
To prove Property~(ii), we first show that
\begin{equation}
\label{eq:L_Y_invariance_proof}
    L_Y \mathcal{F} \subseteq \mathcal{F},\quad \forall Y\in \mathfrak{Y}.
\end{equation}
Define 
\begin{equation}
\label{eq:mathfrak_X_F}
\mathfrak{X}_\mathcal{F} := \{ Y \in \mathfrak{X} (M) \, | \, L_Y \mathcal{F} \subseteq \mathcal{F}\}.
\end{equation}
This is a Lie subalgebra. 
Indeed, for every $Z,Y \in \mathfrak{X}_\mathcal{F}$, $\alpha, \beta \in \mathbb{R}$, and $f \in \mathcal{F}$, we have
\[
\begin{aligned}
L_{\alpha Y + \beta Z} f &= \alpha L_Y f + \beta L_Z f \in \mathcal{F} \\
L_{[Y,Z]} f &= (L_Y L_Z - L_Z L_Y) f \in \mathcal{F}, 
\end{aligned}
\]
which yields $\alpha Y + \beta Z\in \mathfrak{X}_\mathcal{F}$ and $[Y,Z] \in \mathfrak{X}_\mathcal{F}$.
Since we assumed the existence of a Koopman bilinear embedding, \eqref{eq:Koopman_invariant_subspace_control_affine} gives $X_0,\ldots,X_m \in \mathfrak{X}_\mathcal{F}$. 
Moreover, since $\operatorname{Lie}(X_0,\ldots,X_m)$ is the smallest subalgebra containing $X_0,\ldots,X_m$ (recall Definition~\ref{def:generated_lie_algebra}), $X_0,\ldots,X_m \in \mathfrak{X}_\mathcal{F}$ yields $\mathfrak{Y} = \operatorname{Lie}(X_0,\ldots,X_m) \subseteq \mathfrak{X}_\mathcal{F}$. 
This, together with \eqref{eq:mathfrak_X_F}, gives \eqref{eq:L_Y_invariance_proof}. 
Consequently, $\pi_\mathfrak{Y} (Y) = L_Y \in \operatorname{End}(\mathcal{F})$ for every $Y\in \mathfrak{Y}$, so $\pi_\mathfrak{Y}$ is well-defined. 
}

It remains to show that $\pi_\mathfrak{Y}$ is injective. 
If $Y\in \operatorname{Ker}\pi_{\mathfrak{Y}}$, 
\[
L_Y \psi_k (x) = d \psi_k (x) Y(x) = 0,\quad x\in M, \quad k=1,\ldots,N.
\]
Equivalently, $d\Psi (x) Y(x) = 0$ for every $x\in M$. 
Since $\Psi$ is an embedding, $d\Psi (x):T_x M \to T_{\Psi(x)} \mathbb{C}^N$ is injective. 
Thus, $d\Psi (x) Y(x) = 0$ ($\forall x\in M$) implies $Y=0$, and hence $\pi_\mathfrak{Y}$ is injective.

$(\Longleftarrow)$ Assume that there exists a finite-dimensional subspace $\mathcal{F}$ satisfying Properties~(i) and (ii). 
Set $N=\dim \mathcal{F}$ and let $\psi_1,\ldots,\psi_N$ be the basis of $\mathcal{F}$ satisfying Property~(i). 
Since $\pi_\mathfrak{Y}$ is well-defined, $L_Y \mathcal{F} \subseteq \mathcal{F}$ for every $Y \in \mathfrak{Y}$, including $X_0,\ldots ,X_m$. 
Thus, we have $L_{X_j} \mathcal{F} \subseteq \mathcal{F}$. 
Therefore, for every $j=0,\ldots,m$, there exists a matrix $A_j = (a_{k\ell}^{(j)})_{1\leq k,\ell \leq N} \in \mathbb{C}^{N\times N}$ such that 
\[
L_{X_j} \psi_k = \sum_{\ell=1}^N a_{k\ell}^{(j)} \psi_\ell ,\quad k=1,\ldots,N. 
\]
Thus, 
$\Psi := (\psi_1,\ldots,\psi_N)^\top$ gives $L_{X_j} \Psi = A_j \Psi$, and hence $(\Psi,A_j)_{0\leq j\leq m}$ is a Koopman bilinear embedding.
\end{proof}

\section{Proof of Theorem~\ref{thm:ideal_nilpotent}}
\label{sec:proof_ideal_nilpotent}

\begin{proof}
By Condition~\textup{(B)}, there exists $Z\in\mathfrak{I}$ such that the autonomous system $(M,Z)$ admits a finite-algebraic-multiplicity Koopman linear embedding $(\Psi_Z,A_Z)$.
Define
\[
\mathcal{F}_Z := \operatorname{span}_{\mathbb{C}} \left\{ \phi\in C^\infty(M;\mathbb{C}) \;\middle|\; \phi \text{ is a component of }\Psi_Z \right\}.
\]
Then $\sigma(A_Z)=\sigma(L_Z|_{\mathcal{F}_Z})$. 
Moreover, by the finite-algebraic-multiplicity property, the algebraic root space $\mathcal{R}_\lambda(L_Z)$ is finite-dimensional for every $\lambda\in\sigma(A_Z)$, which implies that
\begin{equation}
\label{eq:common_KIS}
\mathcal{F} := \textstyle\bigoplus_{\lambda\in\sigma(A_Z)} \mathcal{R}_\lambda(L_Z)
\end{equation}
is finite-dimensional. 
It also follows that $\mathcal{F}_Z\subseteq\mathcal{F}$.
Indeed, if $\chi_Z(t) = \prod_{\lambda\in\sigma(A_Z)}(t-\lambda)^{m_\lambda}$ is the characteristic polynomial of $L_Z|_{\mathcal F_Z}$, then Cayley--Hamilton theorem gives $\chi_Z(L_Z)\mathcal F_Z=\{0\}$ or, equivalently, $\mathcal F_Z \subseteq \operatorname{Ker} \chi_Z(L_Z)$. 
Lemma~\ref{lem:primary_decomposition} with $V=C^\infty(M;\mathbb{C})$, $T = L_Z$, and $p = \chi_Z$ 
then gives
\[
\begin{aligned}
    \mathcal F_Z &\subseteq \operatorname{Ker} \chi_Z(L_Z) = \textstyle\bigoplus_{\lambda\in\sigma(A_Z)} \operatorname{Ker}(L_Z - \lambda I)^{m_\lambda} \subseteq \mathcal F.
\end{aligned}
\]
In particular, we have $\sigma (L_Z|_{\mathcal{F}}) = \sigma(L_Z|_{\mathcal{F}_Z} ) = \sigma(A_Z)$.

The goal of the proof is to show that
\begin{equation}
\label{eq:common_invariance_full}
L_X\mathcal{F}\subseteq\mathcal{F}, \qquad \forall X\in\mathfrak{Y}.
\end{equation}
Once this has been established, the invariance of $\mathcal{F}$ allows us to construct a $C^\infty$ map $\Psi$ whose components form a basis of $\mathcal{F}$, together with a matrix-valued map\footnote{For a fixed choice of basis for $\mathcal{F}$, the map $A:\mathfrak{Y}\to\mathbb{C}^{N\times N}$ is uniquely determined. Moreover, this is linear because $X\mapsto L_X$ is linear.} $A:\mathfrak{Y} \to \mathbb{C}^{N\times N}$ such that 
\[
L_X\Psi=A(X)\Psi ,\qquad \forall X\in\mathfrak{Y}.
\]
Furthermore, since $\mathcal{F}_Z\subseteq\mathcal{F}$, there exists a matrix $C$ such that $\Psi_Z=C\Psi$. 
Because $\Psi_Z$ is an embedding, $\Psi$ is also an embedding. 
Hence $(\Psi,A (X_j))_{0\leq j\leq m}$ is a Koopman bilinear embedding and the proof is completed. 

We first establish the following invariance property for the nilpotent ideal $\mathfrak{I}$.
\begin{lemma}
\label{lem:nilpotent}
$L_Y\mathcal{F}\subseteq\mathcal{F}$ for every $Y \in \mathfrak{I}$. 
\end{lemma}

\begin{proof}[Proof of Lemma~\ref{lem:nilpotent}]
Since $\mathcal{R}_\lambda(L_Z)$ is finite-dimensional for every $\lambda\in\sigma(A_Z)$, there exists a positive integer $q_\lambda$ such that
\begin{equation}
\label{eq:root_q}
\begin{aligned}
&\operatorname{Ker}(L_Z - \lambda I) \subseteq
\operatorname{Ker}(L_Z - \lambda I)^2 \subseteq\cdots \\ 
&\cdots \subseteq \operatorname{Ker}(L_Z - \lambda I)^{q_\lambda} = \operatorname{Ker}(L_Z-\lambda I)^{q_\lambda+1} = \cdots \\
&\qquad\qquad\qquad\qquad\qquad\qquad\qquad \quad \cdots = \mathcal{R}_\lambda(L_Z).
\end{aligned}
\end{equation}

Choose a positive integer $r$ such that every $(r+1)$-fold iterated Lie bracket of elements of $\mathfrak{I}$ vanishes, as in Definition~\ref{def:nilpotent}.
We claim that, for every $f\in\mathcal{R}_\lambda(L_Z)$ and every $Y\in\mathfrak{I}$,
\begin{equation}
\label{eq:generalized_eigenspace_common}
(L_Z-\lambda I)^{q_\lambda+r-1}L_Yf=0.
\end{equation}
Let $\operatorname{ad}_Z^k$ and $\operatorname{ad}_{L_Z}^k$ denote the $k$-fold adjoint actions of $Z$ and $L_Z$, respectively. 
Since $L_{[Z,Y]}=[L_Z,L_Y]_{\operatorname{End}}$, it follows that $\operatorname{ad}_{L_Z}^k(L_Y)=L_{\operatorname{ad}_Z^kY}$ for every $k\geq 0$. 
The nilpotency of $\mathfrak I$ therefore implies that $\operatorname{ad}_{L_Z}^k L_Y = 0$ for $k\geq r$. 
Applying Lemma~\ref{lem:app} with $V=C^\infty(M;\mathbb C)$, $A=L_Z-\lambda I$, $B=L_Y$, and $n=q_\lambda+r-1$, we obtain
\begin{equation}
\label{eq:binomial_2}
\begin{aligned}
&(L_Z-\lambda I)^{q_\lambda+r-1}L_Y \\ &=
\sum_{k=0}^{r-1} {\textstyle\binom{q_\lambda+r-1}{k}} \pi (\operatorname{ad}_{L_Z}^k{L_Y}) (L_Z-\lambda I)^{q_\lambda+r-1-k},
\end{aligned}
\end{equation}
where we used $\operatorname{ad}_{L_Z-\lambda I}=\operatorname{ad}_{L_Z}$ and $\operatorname{ad}_{L_Z}^k L_Y = 0$ for $k\geq r$. 
Then, for every $f\in\mathcal R_\lambda(L_Z)=\operatorname{Ker}(L_Z-\lambda I)^{q_\lambda}$, every term on the right-hand side of \eqref{eq:binomial_2} annihilates $f$.
Thus, applying \eqref{eq:binomial_2} to $f$ yields \eqref{eq:generalized_eigenspace_common}.

Equations \eqref{eq:generalized_eigenspace_common} and \eqref{eq:root_q} imply that $L_Yf\in\mathcal{R}_\lambda(L_Z)$ for every $f\in\mathcal{R}_\lambda(L_Z)$.
This holds for every $\lambda\in\sigma(A_Z)$.
Consequently, the space $\mathcal{F}$ defined in \eqref{eq:common_KIS} satisfies $L_Y\mathcal{F}\subseteq\mathcal{F}$, which proves the lemma. 
\end{proof}

Next, consider the characteristic polynomial
\[
\chi(t) := \det\bigl(tI_{\mathcal{F}}-L_Z|_{\mathcal{F}}\bigr)
\]
of the linear operator $L_Z|_{\mathcal{F}}$.
Since $\sigma(L_Z|_{\mathcal{F}})=\sigma(A_Z)$, there exist positive integers $d_\lambda$ with $\lambda\in\sigma(A_Z)$ such that
\begin{equation}
\label{eq:determinant}
\chi(t) = \textstyle \prod_{\lambda \in \sigma(A_Z)}  (t-\lambda)^{d_\lambda}.
\end{equation}
Applying the Cayley--Hamilton theorem to $L_Z|_\mathcal{F}$ gives
\begin{equation}
\label{eq:cayley_hamilton}
\chi(L_Z)f=0, \qquad \forall f\in\mathcal{F}.
\end{equation}

Since $Z\in\mathfrak{I}$, where $\mathfrak{I}$ is an ideal, Lemma~\ref{lem:nilpotent} yields the following result.

\begin{lemma}
\label{lem:ideal}
$[\chi(L_Z),L_X]_{\mathrm{End}}\mathcal{F}
\subseteq
\mathcal{F}$ for every $X\in\mathfrak{Y}$.
\end{lemma}

\begin{proof}[Proof of Lemma~\ref{lem:ideal}]
Since $\mathfrak{I}$ is an ideal of $\mathfrak{Y}$, $Z\in\mathfrak{I}$ implies $[Z,X]\in\mathfrak{I}$.
It follows from Lemma~\ref{lem:nilpotent} that $[L_Z,L_X]_{\mathrm{End}}\mathcal{F} = L_{[Z,X]}\mathcal{F} \subseteq \mathcal{F}$. 
Moreover, for every positive integer $k$,
\[
\begin{aligned}
&[L_Z^{k+1},L_X]_{\mathrm{End}} = L_Z^{k+1}L_X-L_XL_Z^{k+1} \\
&= \sum_{\ell=0}^{k} L_Z^\ell(L_ZL_X-L_XL_Z)L_Z^{k-\ell} = \sum_{\ell=0}^{k} L_Z^\ell L_{[Z,X]}L_Z^{k-\ell}.
\end{aligned}
\]
Since both $L_Z$ and $L_{[Z,X]}$ preserve $\mathcal{F}$, we have $[L_Z^{k+1},L_X]_{\mathrm{End}} \mathcal{F} \subseteq \mathcal{F}$.
Since $\chi$ is a polynomial, we conclude that $[\chi(L_Z),L_X]_{\mathrm{End}} \mathcal{F} \subseteq\mathcal{F}$, which proves the lemma.
\end{proof}

We now prove \eqref{eq:common_invariance_full}.
Let $f\in\mathcal{F}$ and $X\in\mathfrak{Y}$.
The definition of $[\cdot,\cdot]_{\mathrm{End}}$, Equation~\eqref{eq:cayley_hamilton}, and Lemma~\ref{lem:ideal} give
\[
\begin{aligned}
\chi(L_Z)L_Xf &= L_X\, \chi(L_Z)f + [\chi(L_Z),L_X]_{\mathrm{End}}f\\
&= [\chi(L_Z),L_X]_{\mathrm{End}}f \in \mathcal{F}.
\end{aligned}
\]
Applying $\chi(L_Z)$ again to this identity and using \eqref{eq:cayley_hamilton} yields
$\chi(L_Z)^2 L_X f=0$, and hence $L_Xf\in \operatorname{Ker}\chi(L_Z)^2$. 
Then, Lemma~\ref{lem:primary_decomposition} with $V=C^\infty(M;\mathbb{C})$, $T=L_Z$, and $p=\chi^2$ gives 
\[
\begin{aligned}
\operatorname{Ker}\chi(L_Z)^2 &= \textstyle\bigoplus_{\lambda\in\sigma(A_Z)} \operatorname{Ker}(L_Z-\lambda I)^{2d_\lambda} \\
&\subseteq \textstyle\bigoplus_{\lambda\in\sigma(A_Z)}  \mathcal{R}_\lambda(L_Z) = \mathcal{F},
\end{aligned}
\]
where the final equality follows from \eqref{eq:common_KIS}. 
We thus obtain $L_X f \in \operatorname{Ker}\chi(L_Z)^2 \subseteq \mathcal{F}$. Therefore, \eqref{eq:common_invariance_full} follows and the proof is completed.
\end{proof}

\section{Collection of Lemmas}

\begin{lemma}
\label{lem:primary_decomposition}
Let $V$ be a possibly infinite-dimensional complex vector space and let $T:V\to V$ be a linear operator.
Suppose that $p(t)=\prod_{j=1}^s p_j(t)$ where $p_1,\ldots,p_s$ are complex polynomials that are pairwise relatively prime. 
Then
\[
\operatorname{Ker}p(T) = \textstyle \bigoplus_{j=1}^s \operatorname{Ker}p_j(T).
\]
\end{lemma}

\begin{proof}
See \cite[Theorem~2.2.14]{Brady2000}. 
\end{proof}

\begin{lemma}
\label{lem:app}
Let $V$ be a vector space and let $A,B\in\operatorname{End}(V)$. 
Then, for any integer $n\geq 0$,
\[
A^n B = \sum_{k=0}^n {\textstyle \binom{n}{k}} \operatorname{ad}_A^k (B) A^{n-k}.
\]
\end{lemma}

\begin{proof}
Clearly, $A^iC = (A^{i-1}C)A + A^{i-1}\operatorname{ad}_AC$ for every $C\in\operatorname{End}(V)$ and $i\geq 1$. 
Applying this formula to $A^n B$ by replacing $C=B,\operatorname{ad}_A B, \operatorname{ad}_A^2 B,\ldots$ gives the result.
\end{proof}

\bibliographystyle{IEEEtran}
\bibliography{Refs}

\end{document}